\documentclass{sig-alternate}

\usepackage[usenames,dvipsnames]{color}
\usepackage{amssymb}
\usepackage{multirow}
\usepackage{hyperref}
\usepackage{graphicx}

\DeclareGraphicsExtensions{.pdf}
  
\newcommand{\secref}[1]{Section~\ref{sec_#1}}
\newcommand{\figref}[1]{Figure~\ref{fig_#1}}
\newcommand{\tblref}[1]{Table~\ref{tbl_#1}}

\newcommand{\ler}[0]{l_{\mathrm{ER}}}
\newcommand{\cer}[0]{C_{\mathrm{ER}}}

\newcommand{\dcent}[0]{{\mathrm{DC}}}
\newcommand{\ccent}[0]{{\mathrm{CC}}}
\newcommand{\bcent}[0]{{\mathrm{BC}}}
\newcommand{\drive}[0]{n_{\mathrm{d}}/n}

\newcommand{\mo}[0]{MO\space}
\newcommand{\cp}[0]{CP\space}
\newcommand{\mm}[0]{MM\space}
\newcommand{\gp}[0]{GP\space}

\newcommand{\flm}[0]{\textit{flmng}\space}

\newcommand{\clt}[0]{\textit{colt}\space}
\newcommand{\jng}[0]{\textit{jung}\space}

\newcommand{\org}[0]{\textit{org}\space}
\newcommand{\wek}[0]{\textit{weka}\space}
\newcommand{\jvx}[0]{\textit{javax}\space}
\newcommand{\jav}[0]{\textit{java}\space}

\newcommand{\weks}[0]{\textit{weka}}

\definecolor{red}{RGB}{147, 82, 0}
\definecolor{green}{RGB}{146, 144, 0}
\definecolor{blue}{RGB}{0, 84, 147}
\definecolor{gray}{RGB}{121, 121, 121}

\begin{document}


\title{Software Systems through Complex Networks Science: Review, Analysis and Applications}

\numberofauthors{2}
\author{
\alignauthor
Lovro \v{S}ubelj\\
       \affaddr{University of Ljubljana}\\
	\affaddr{Faculty of Computer and Information Science}\\
       \affaddr{Tr\v{z}a\v{s}ka cesta 25, SI-1000 Ljubljana, Slovenia}\\
       \email{lovro.subelj@fri.uni-lj.si}
\alignauthor
Marko Bajec\\
       \affaddr{University of Ljubljana}\\
	\affaddr{Faculty of Computer and Information Science}\\
       \affaddr{Tr\v{z}a\v{s}ka cesta 25, SI-1000 Ljubljana, Slovenia}\\
       \email{marko.bajec@fri.uni-lj.si}
}

\date{10 June 2012}

\conferenceinfo{SoftwareMining '12}{, August 12, Beijing, China}
\CopyrightYear{2012}
\crdata{978-1-4503-1560-9/12/08} 

\maketitle


\begin{abstract}
Complex software systems are among most sophisticated human-made systems, yet only little is known about the actual structure of 'good' software. We here study different software systems developed in Java from the perspective of network science. The study reveals that network theory can provide a prominent set of techniques for the exploratory analysis of large complex software system. We further identify several applications in software engineering, and propose different network-based quality indicators that address software design, efficiency, reusability, vulnerability, controllability and other. We also highlight various interesting findings, e.g., software systems are highly vulnerable to processes like bug propagation, however, they are not easily controllable.
\end{abstract}






\section{Introduction}
\label{sec_intro}
Complex software systems are among most sophisticated systems ever created by human. Nevertheless, only little is known about the actual structure and quantitative properties of large software systems~\cite{CY09}. For instance, in the context of software engineering, one is interested in how 'good' software looks like. Commonly adopted approaches and techniques fail to give a comprehensive answer~\cite{Bei90,CGP00}, moreover, there is also a lack of a simple but yet rigorous framework for software analysis (to our knowledge). The above dilemma was denoted \textit{software law} problem~\cite{CY09}, which urges towards identifying (physical) laws obeyed by software systems that could be used in practical applications.

Networks possibly provide the most adequate framework for the analysis of the structure of complex systems like software projects\footnote{Throughout the paper, the term project refers to a repository of software code.}. Also, due to their simple and intelligible form, analysis of different networks has already provided several significant discoveries in the last decade~\cite{WS98,BA99,GN02,LSB11b}. Note that the adoption of software networks is not novel~\cite{VCS02,Mye03,Koh09,SB11s}, however, network analysis is still only rarely used in software engineering. The main purpose of this study is thus to highlight different techniques developed in the field of network analysis, and to expose their use in software comprehension, development and engineering. We review most of the past work on different types of software networks, whereas we also include network analysis techniques proposed just recently~\cite{LSB11b,SB12u}. (Note that the main focus of the paper is merely a review, rather than a detailed comparison of network analysis techniques with other approaches.)

The study in the paper analyses software networks on different levels of granularity. First, we address the macroscopic properties of software networks like scale-free and small-world phenomena~\cite{WS98,BA99} that are related to the structure and design of the entire project, or projects, represented by the network. Second, we analyze the microscopic properties of individual nodes, with special emphasis on different dynamical processes occurring on software networks like bug propagation~\cite{AJB00,PV01}. The above can be related to software quality, complexity, reusability, robustness, vulnerability and controllability. Third, we also identify mesoscopic structural modules within software networks~\cite{GN02,SB12u} and show their applicability in the context of software abstraction and refactoring. The paper thus exposes network analysis as a prominent set of techniques for software engineering.

The rest of the paper is structured as follows. \secref{nets} introduces software networks used in the study. \secref{ana} analyzes different characteristics of adopted networks and discusses their use in software engineering. Some applications of the presented techniques are given in~\secref{appl}, while \secref{conc} concludes the paper.


\begin{figure*}[t]
\centering
\includegraphics[width=2.00\columnwidth]{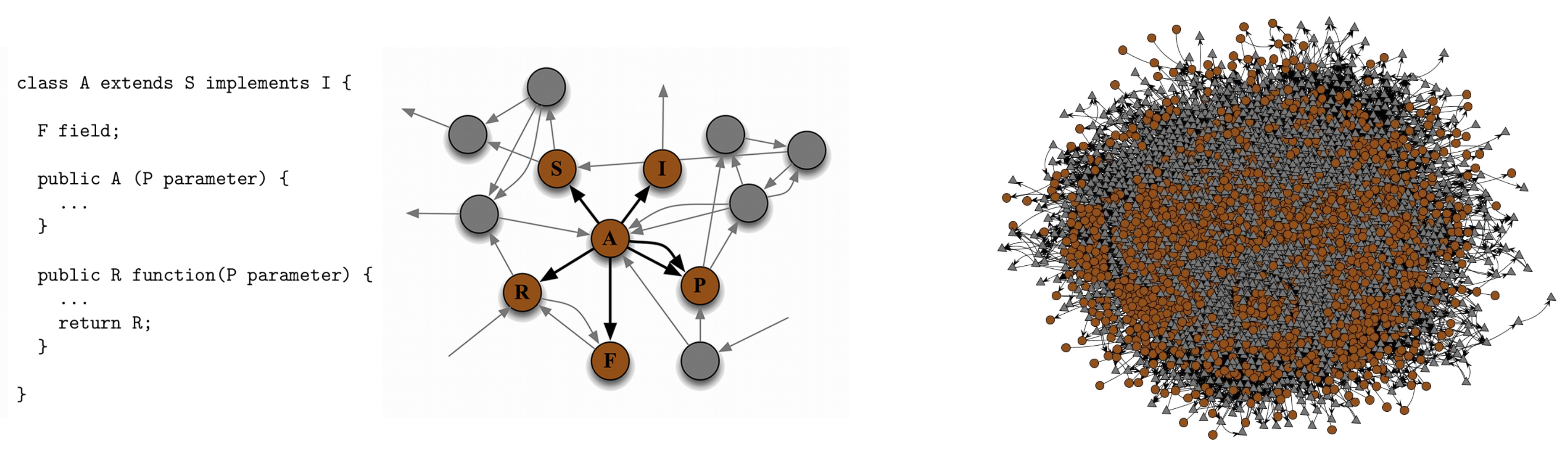}
\caption{\label{fig_toy}(left)~A simple Java class and the corresponding part of class dependency network. Direction of links is (mostly) just the opposite to the flow of information. (right)~Class dependency network of \texttt{java} ({\color{red}circles}) and \texttt{javax} ({\color{gray}triangles}) namespaces of Java language.}
\end{figure*}

\begin{table*}[t]
\centering
\caption{\label{tbl_nets}Properties of class dependency networks used in the study.}
\begin{tabular}{|c|l|r|r|r|r|r|r|}
\hline
\multicolumn{1}{|c}{Network} & 
\multicolumn{1}{|c}{Project} & 
\multicolumn{1}{|c}{$n$} & 
\multicolumn{1}{|c}{$m$} & 
\multicolumn{1}{|c}{$k$} & 
\multicolumn{1}{|c}{LCC} &
\multicolumn{1}{|c}{$|A|$} & 
\multicolumn{1}{|c|}{$|P|$}\\\hline
\flm & Flamingo 4.1 (GUI components)~\cite{SB11s} & $141$ & $269$ & $3.82$ & $0.88$ & $153$ & $18$ \\\hline
\clt & Colt 1.2.0 (scientific computation)~\cite{SB11s} & $243$ & $720$ & $5.93$ & $0.94$ & $267$ & $21$ \\\hline
\jng & JUNG 2.0.1 (network analysis)~\cite{SB11s} & $317$ & $719$ & $4.54$ & $0.96$ & $357$ & $41$ \\\hline
\org & Java 1.6.0.7 (\texttt{org} namespace)~\cite{SB11s} & $709$ & $3571$ & $10.07$ & $0.69$ & $778$ & $50$ \\\hline
\wek & Weka 3.6.6 (data mining framework) & $953$ & $4097$ & $8.60$ & $0.98$ & $1054$ & $84$ \\\hline
\jvx & Java 1.6.0.7 (\texttt{javax} namespace)~\cite{SB11s} & $1595$ & $5287$ & $6.63$ & $0.44$ & $1889$ & $118$ \\\hline
\jav & Java 1.6.0.7 (\texttt{java} namespace)~\cite{SB11s} & $1516$ & $10049$ & $13.26$ & $1.00$ & $1518$ & $56$ \\\hline
\end{tabular}
\end{table*}

\section{Software networks}
\label{sec_nets}
Various types of networks have been proposed for the analysis of the structure of complex software systems. For instance, software architecture maps~\cite{VCS02}, software mirror graphs~\cite{CY09}, class, method and package collaboration graphs~\cite{HCK06}, subrutine call graphs~\cite{Mye03}, inter-package dependency networks~\cite{LW04}, software class diagrams~\cite{VS05b} and class dependency networks~\cite{SB11s}, to name just a few. Networks mainly divide whether they are constructed from source code, byte code or program execution traces, and due to the level of software architecture represented by the nodes, and the set of interdependencies represented by the links. 

For consistency with some previous work~\cite{BFNRSVMT06,HCK06,WKD07,SB11s}, we construct networks from the source code of different Java projects\footnote{Networks are available from \href{http://lovro.lpt.fri.uni-lj.si/}{http://lovro.lpt.fri.uni-lj.si/}.} (\tblref{nets}). Due to the object-oriented view of Java language, nodes in the network can represent either project packages, software classes, methods and functions or individual lines of code. We here adopt class dependency networks~\cite{SB11s}, where nodes represent classes and links correspond to different dependencies among them (\figref{toy}). The latter is based on the following reasons. First, as networks are constructed merely from the signatures of different classes, and functions and fields therein, they are only mildly influenced by the subjective nature of each individual developer. (This can be more adequately modeled by, e.g., text mining applied to the names of different programming constructs~\cite{KDG07}.) Second, mesoscopic structures of class dependency networks coincide with project packages~\cite{SB11s,SB12u}, which enables various applications in software engineering (\secref{appl}). Third, such networks relate to the information flow between different parts of software project, and also coincide with the human comprehension of the object-oriented software systems.

Note that class dependency networks address only the inter-class structure of the software project, whereas the intra-class dependencies are disregarded. However, similarly as above, the latter reflect also the programming style of a particular developer, rather than the intrinsic structure of the software project alone. Nevertheless, future work will extend the study to inter- and intra-class dependencies using to concepts of interdependent or coupled networks~\cite{MRMPO10,GBSH12}.

\begin{figure*}[t]
\centering
\includegraphics[width=2.00\columnwidth]{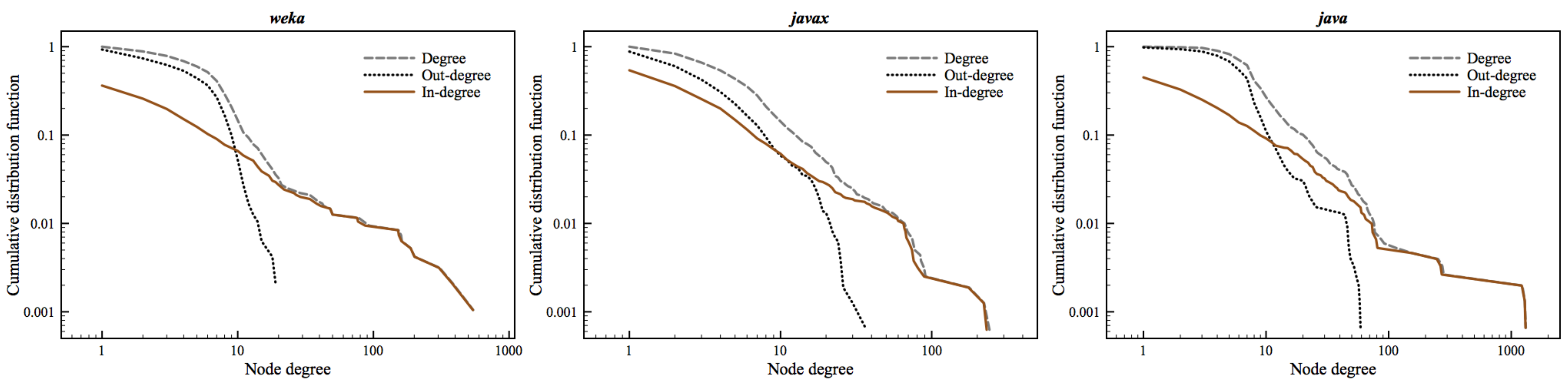}
\caption{\label{fig_scale}Degree distributions of \weks, \jvx and \jav networks.}
\end{figure*}

\begin{table*}[t]
\centering
\caption{\label{tbl_world}Different statistics for class dependency networks used in the study.}
\begin{tabular}{|c|r|r|r|r|r|r|r|r|}
\hline
\multicolumn{1}{|c}{\rule{0pt}{12pt}Network} & 
\multicolumn{1}{|c}{$\gamma$} & 
\multicolumn{1}{|c}{$C$} & 
\multicolumn{1}{|c}{$D$} & 
\multicolumn{1}{|c}{$\cer$} & 
\multicolumn{1}{|c}{$l$} &
\multicolumn{1}{|c}{$E$} &
\multicolumn{1}{|c}{$\ler$} &
\multicolumn{1}{|c|}{$\drive$}\\\hline
\flm & $3.0$ & $0.25$ & $0.31$ & $\mathit{0.03}$ & $4.05$ & $0.03$ & $\mathit{3.47}$ & $0.38$ \\\hline
\clt & $2.7$ & $0.41$ & $0.47$ & $\mathit{0.02}$ & $3.44$ & $0.03$ & $\mathit{3.16}$ & $0.30$ \\\hline
\jng & $2.5$ & $0.37$ & $0.42$ & $\mathit{0.01}$ & $4.19$ & $0.02$ & $\mathit{3.88}$ & $0.48$ \\\hline
\org & $2.2$ & $0.57$ & $0.62$ & $\mathit{0.01}$ & $2.68$ & $0.03$ & $\mathit{2.81}$ & $0.39$ \\\hline
\wek & $3.0$ & $0.39$ & $0.43$ & $\mathit{0.01}$ & $2.91$ & $0.01$ & $\mathit{3.39}$ & $0.12$ \\\hline
\jvx & $2.6$ & $0.38$ & $0.44$ & $\mathit{0.00}$ & $3.88$ & $0.02$ & $\mathit{3.16}$ & $0.30$ \\\hline
\jav & $2.4$ & $0.69$ & $0.73$ & $\mathit{0.01}$ & $2.18$ & $0.02$ & $\mathit{3.09}$ & $0.17$ \\\hline 
\end{tabular}
\end{table*}

Formally, let a project consist of classes $A=\{A_1,A_2,\dots\}$ and let $P$ be the set of software packages (bottom-most level of the package hierarchy). Corresponding class dependency network is then a directed graph $G(N,L)$, with nodes $N=\{1,\dots,n\}$ and links $L$ ($m=|L|$). Node $i$ corresponds to a class $A_i$, however, since isolated nodes are discarded in the analysis, $n\leq |A|$. A directed link $(i,j)\in L$ represent some dependency between classes $A_i$ and $A_j$: inheritance ($A_i$ inherits or implements $A_j$), parameter ($A_i$ contains a method, function or constructor that takes $A_j$ as parameter), return  ($A_i$ contains a method or function that returns $A_j$) and field ($A_i$ contains a field of type $A_j$). Denote $k$ to be the average degree in the network (i.e., average number of links incident to a node). Furthermore, let $k^{in}$ and $k^{out}$ be the average in-degree and out-degree of the nodes, $k=k^{in}+k^{out}$. Hence, $k_i^{out}$ corresponds to a number of other classes required to implement the functionality of a respective class $A_i$, while $k_i^{in}$ corresponds to the number of classes that use (depend on) $A_i$. Last, denote LCC to be the fraction of nodes in the largest connected component\footnote{All networks in figures are reduced to LCC-s.}.

\tblref{nets} shows properties of class dependency networks used in the study. Networks were selected thus to represent a diverse set of software systems including utility libraries (e.g., \flm and \clt networks), complete frameworks (e.g., \jng and \wek networks) and also the core of Java language itself (i.e., \jav network).

Software networks are compared against Erd\"{o}s-R\'{e}nyi random graphs~\cite{ER59}, where a link are placed between each pair of $n$ nodes with probability $k/(n-1)$, where $k=2m/n$ for some $n$ and $m$.


\section{Analysis and discussion}
\label{sec_ana}


\subsection{Scale-free networks -- software complexity and reusability}
\label{sec_ana_scale}
Simple random graphs experience a Poisson degree distribution $p_k$, $p_k\sim \frac{\lambda^ke^{-\lambda}}{k!}$. On the contrary, $p_k$ of most real-world networks including software networks follows a power-law form $p_k\sim k^{-\gamma}$~\cite{BA99,VCS02,PNFB05,CMPS07}, where $\gamma$ is a scale-free exponent, $\gamma>1$. The latter can be clearly observed by a straight line with slope $-\gamma$ in a log-log plot (\figref{scale}). Networks with power-law degree distributions are denoted scale-free, while $\gamma$ can be directly related to the spreading processes occurring on networks~\cite{Sin11} (e.g., bug propagation). For $\gamma\in(2,3)$, even a very small fraction of faulty nodes can already render the entire system inapplicable~\cite{PV01,Sin11}. Unfortunately, the latter applies for all software networks used here (\tblref{world}).

Scale-free networks are usually considered an artifact of Yule's process or \textit{rich get richer} phenomena~\cite{BA99}. For class dependency networks, this refers to the fact that highly used classes are, obviously, well known among developers, and would thus also be more commonly adopted in the future. However, power-laws should thus emerge merely in the in-degree distribution $p_k^{in}$ that refers to the number of times each class is used~\cite{VS05a,BFNRSVMT06} (\figref{scale}). More precisely, scale-free nature of $p_k^{in}$ is a result of high code reusability. On the other hand, out-degree distribution $p_k^{out}$ is related to software complexity, since classes with high $k_i^{out}$ encompass most complex functionality. Here, complexity refers to the number of other classes needed to implement the functionality of the respective class. For example, most commonly reused class in \jav network is \texttt{String}, whereas \texttt{FileDialog} is the most complex one (\tblref{hubs}).

Well developed software project should thus exhibit scale-free $p_k^{in}$ and highly truncated $p_k^{out}$. Next, lower $\gamma$ indicates higher code reuse, which also decreases the probability of fault propagation throughout the system. Last, classes with very high $k_i^{out}$, and also $k_i^{in}$, should be implemented with extra care (see~\secref{ana_node}).


\begin{figure*}[t]
\centering
\includegraphics[width=2.00\columnwidth]{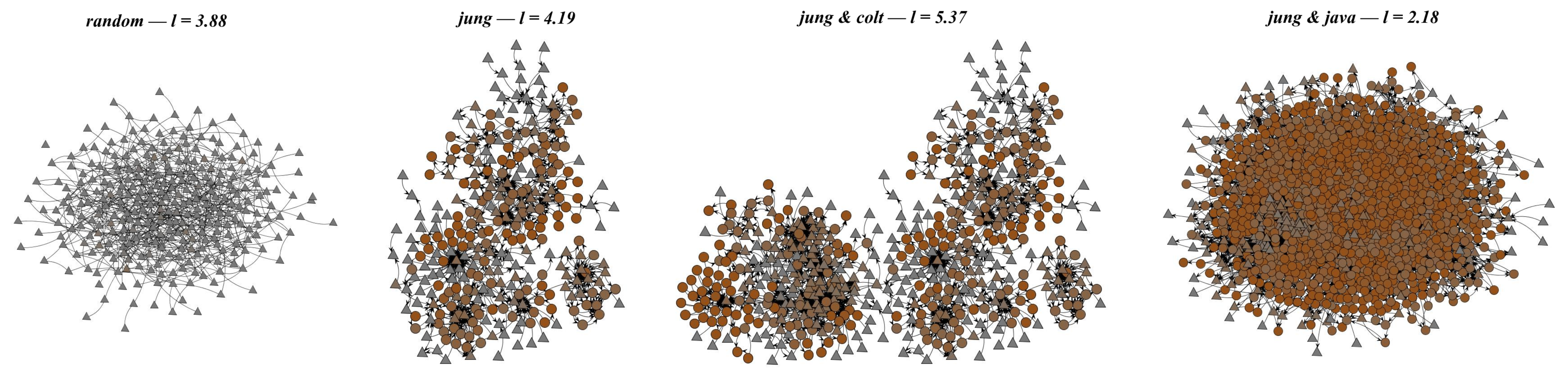}
\caption{\label{fig_jung}A random graph, \jng network, \jng \& \clt network and \jng \& \jav network. Average distance between the nodes $l$ equals $3.88$, $4.19$, $5.37$ and $2.18$. Node symbols correspond to clustering $D$~\cite{SV05} that ranges between $0$ ({\color{gray}triangles}) and $1$ ({\color{red}circles}).}
\end{figure*}

\begin{figure*}[t]
\centering
\includegraphics[width=2.00\columnwidth]{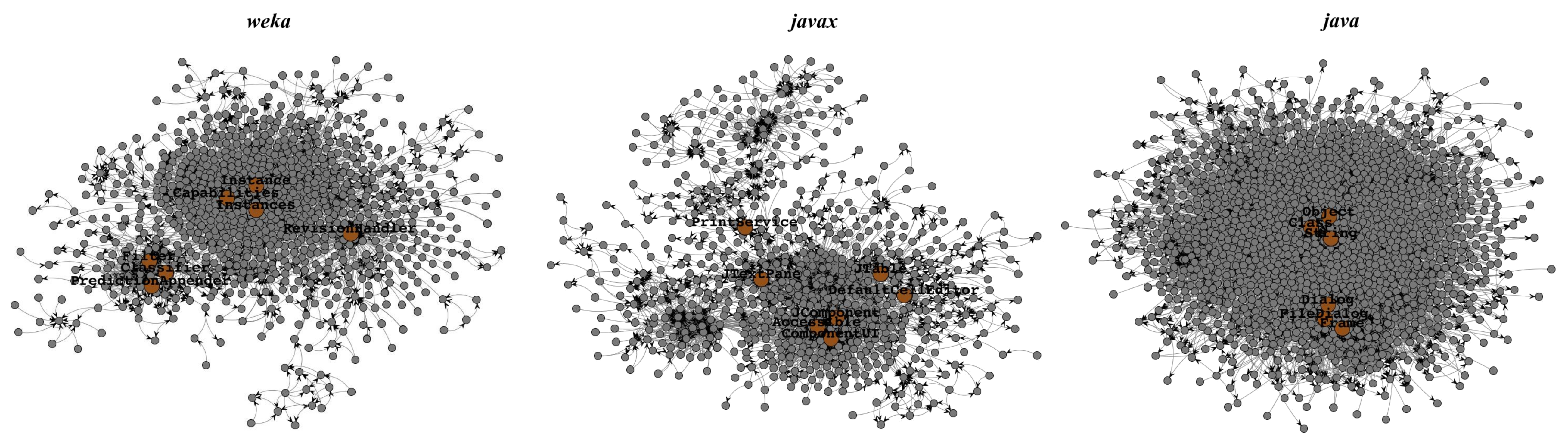}
\caption{\label{fig_seed}\weks, \jvx and \jav networks with highlighted seed nodes.}
\end{figure*}

\subsection{Small-world networks -- software structure and design}
\label{sec_ana_world}
Software networks exhibit small-world phenomena~\cite{WS98} (see \cite{Mye03,VS07} and \tblref{world}), which refers to high clustering $C$~\cite{WS98} and very short average distance between the nodes $l$~\cite{AMO93} (also known as six degrees of separation~\cite{Mil67}). $C$ measures transitivity in the network, and is defined as the probability that two neighbors of a node are also linked, $C\in[0,1]$. $l=\frac{1}{n(n-1)}\sum_{i\neq j}d_{ij}$, where $d_{ij}$ is the distance between $i$ in $j$ in the respective undirected network (i.e., number of links in the shortest path). Small-world networks most commonly refer to $C\gg C_{ER}$ and $l\approx l_{ER}$~\cite{WS98}, where $C_{ER}$ and $l_{ER}$ are the values for a corresponding random graph.

Clustering of software networks can be related to intrinsic characteristics of the underlying systems~\cite{SB12f}. For instance, visualization classes usually experience very high clustering, while clustering is almost zero for I/O classes~\cite{SB12f,SB11g}. 

Average distance $l$ is an important indicator of the structural design of the project, or projects, represented by the network. More precisely, since $l\approx l_{ER}$, $l\gg l_{ER}$ indicates that the underlying software system has divided into several independent parts with rather different functionality (\figref{jung}). Note also that software networks should never be combined with the core of the language, since the latter completely obscures its structure and dynamics.

It ought to be mentioned that software networks are small-world only in the undirected case~\cite{Koh09}. The contrary would imply a cyclic flow of information within the software project. For instance, high-level Java class \texttt{String} does not use the functionality of a lower-level \texttt{FileDialog}. Let $E$ be the efficiency of network information flow~\cite{Lm01} defined as $E=\frac{1}{n(n-1)}\sum_{i\neq j}1/d'_{ij}$, where $d'_{ij}$ is the distance from $i$ to $j$ in a (directed) network, $E\in[0,1]$. Small-worlds should result in high flow efficiency $E$, however, software networks have $E\approx 0$ (\tblref{world}).

Well designed software project should thus experience $C\gg C_{ER}$, $l\approx l_{ER}$ and $E\approx 0$. Also, one should be wary of $l\gg l_{ER}$ throughout the project development. 


\subsection{Network nodes -- software vulnerability and control}
\label{sec_ana_node}
In the context of spreading processes on software networks~\cite{NGS08,WL10a} (e.g., bug propagation) and network robustness~\cite{AJB00,Sin11} (i.e., software vulnerability), one is interested into so called seed nodes that could originate the propagation of faults through the entire system\footnote{Although a poor implementation of any software class already makes the system vulnerable, the problem is even amplified in the case of, e.g., highly reused classes.}. Centrality metrics that measure nodes influence are commonly regarded as a prominent indicator of seed nodes~\cite{Fre77,Fre79}. Denote $\dcent_i$ to be the degree centrality defined as $\dcent_i=k_i/(n-1)$, where $k_i$ is the degree of node $i$, $\dcent_i\in[0,1]$. Next, denote $\ccent_i$ to be the harmonic closeness centrality defined as the average inverse of distance from $i$ to the rest of the nodes, $\ccent_i=\frac{1}{n-1}\sum_{i\neq j}1/d'_{ij}$, $\ccent_i\in[0,1]$. Last, denote $\bcent_i$ to be the betweenness centrality defined as the fraction of shortest paths between the nodes that go through $i$, $\bcent_i\in[0,1]$.

\begin{table*}[t]
\centering
\caption{\label{tbl_hubs}Hubs (i.e., nodes with very high degree) within \weks, \jvx and \jav networks.}
\begin{tabular}{|c|r|r|c|r|r|c|r|r|}
\hline
\multicolumn{3}{|c}{\wek} & 
\multicolumn{3}{|c}{$\jvx$} & 
\multicolumn{3}{|c|}{$\jav$}\\\hline
\multicolumn{1}{|c}{Node} & 
\multicolumn{1}{|c}{$k_i^{in}$} & 
\multicolumn{1}{|c}{$k_i^{out}$} & 
\multicolumn{1}{|c}{Node} & 
\multicolumn{1}{|c}{$k_i^{in}$} & 
\multicolumn{1}{|c}{$k_i^{out}$} & 
\multicolumn{1}{|c}{Node} & 
\multicolumn{1}{|c}{$k_i^{in}$} & 
\multicolumn{1}{|c|}{$k_i^{out}$}\\\hline
\texttt{Instances} & $541$ & $5$ & \texttt{JComponent} & $235$ & $11$ & \texttt{String} & $1308$ & $7$ \\\hline
\texttt{Instance} & $381$ & $4$ & \texttt{Accessible} & $222$ & $1$ & \texttt{Class} & $1288$ & $4$ \\\hline
\texttt{Capabilities} & $304$ & $4$ & \texttt{ComponentUI} & $175$ & $2$ & \texttt{Object} & $1228$ & $1$ \\\hline
\texttt{ClassAssigner} & $0$ & $19$ & \texttt{JTable} & $6$ & $37$ & \texttt{FileDialog} & $0$ & $59$ \\\hline
\texttt{Filter} & $0$ & $19$ & \texttt{JTextPane} & $0$ & $30$ & \texttt{Frame} & $4$ & $58$ \\\hline
\texttt{Classifier} & $0$ & $18$ & \texttt{JMenu} & $1$ & $26$ & \texttt{Dialog} & $5$ & $57$ \\\hline
\end{tabular}
\end{table*}

\begin{table*}[t]
\centering
\caption{\label{tbl_seed}Seed nodes (i.e., very influential nodes) within \weks, \jvx and \jav networks.}
\begin{tabular}{|c|r|r|c|r|r|c|r|r|}
\hline
\multicolumn{3}{|c}{\wek} & 
\multicolumn{3}{|c}{\jvx} & 
\multicolumn{3}{|c|}{\jav}\\\hline
\multicolumn{1}{|c}{\rule{0pt}{12pt}Node} & 
\multicolumn{1}{|c}{$\ccent_i$} & 
\multicolumn{1}{|c}{$\bcent_i$} & 
\multicolumn{1}{|c}{\rule{0pt}{12pt}Node} & 
\multicolumn{1}{|c}{$\ccent_i$} & 
\multicolumn{1}{|c}{$\bcent_i$} & 
\multicolumn{1}{|c}{\rule{0pt}{12pt}Node} & 
\multicolumn{1}{|c}{$\ccent_i$} & 
\multicolumn{1}{|c|}{$\bcent_i$}\\\hline
\texttt{PredictionAppender} & $0.03$ & $0.00$ & \texttt{DefaultCellEditor} & $0.10$ & $0.00$ & \texttt{FileDialog} & $0.09$ & $0.00$ \\\hline
\texttt{Classifier} & $0.03$ & $0.01$ & \texttt{JTable} & $0.10$ & $0.12$ & \texttt{Dialog} & $0.09$ & $0.00$ \\\hline
\texttt{Filter} & $0.03$ & $0.00$ & \texttt{JTextPane} & $0.09$ & $0.08$ & \texttt{Frame} & $0.09$ & $0.00$ \\\hline
\texttt{Instances} & $0.01$ & $0.51$ & \texttt{JComponent} & $0.04$ & $0.23$ & \texttt{String} & $0.02$ & $0.36$ \\\hline
\texttt{RevisionHandler} & $0.00$ & $0.26$ & \texttt{Accessible} & $0.01$ & $0.18$ & \texttt{Object} & $0.02$ & $0.32$ \\\hline
\texttt{Instance} & $0.01$ & $0.13$ & \texttt{PrintService} & $0.02$ & $0.17$ & \texttt{Class} & $0.02$ & $0.26$ \\\hline
\end{tabular}
\end{table*}

As $k_i\approx k_i^{in}$ for software networks, $\dcent_i$ actually identifies classes with the highest code reuse or, equivalently, high in-degree $k_i^{in}$ (\tblref{hubs}). Similar set of influential classes is reported by~$\bcent_i$ (\tblref{seed}). On the other hand, $\ccent_i$ identifies classes that somewhat coincide with high complexity classes identified in~\secref{ana_scale}.  $\bcent_i$ (and $\dcent_i$) thus reveals classes whose faulty implementation could influence the entire system, whereas $\ccent_i$ exposes classes that are most prone to an arbitrary fault within the system. The former commonly reside in the core of the respective software network, while the latter are found in the periphery~(\figref{seed}).

Extra care should be put in the development of classes with high $\bcent_i$, while high $\ccent_i$ classes can be adopted for an effective, and also efficient, software testing.

Network controllability has just recently been proposed for the analysis of directed real-world networks~\cite{LH07,LSB11b}. Here, one is particularly interested in the number of driver nodes $n_d$ that one has to govern in order guide the entire system~\cite{LSB11b} (i.e., gain control over the output of the system under the assumption of simple linear transformations). For scale-free networks with $p_k^{in}$ equal to $p_k^{out}$, $n_d/n\approx e^{k(\gamma-2)/(2-2\gamma)}$, $\gamma>2$~\cite{LSB11b}. Note that, contrary to seed nodes (\tblref{seed}) and general belief, driver nodes tend to avoid high degree nodes~\cite{LSB11b,Ege11a}.

Most software network are not highly controllable, since one would have to manage $30$-$50\%$ of classes in order to control the entire project (\tblref{world}). Nevertheless, due to high density, the core of Java language can be controlled through merely $17\%$ of classes in \textit{java} namespace. For comparison, $n_d/n$ equals $\approx 80\%$ for regulatory networks, $\approx 50\%$ for the Internet, $\approx 30\%$ for power grids and on-line social networks, while, interestingly, it is below $3\%$ for corporate ownership networks~\cite{LSB11b}.

Controllability of a software system can be limited by decreasing $k$ or $\gamma$, which is achieved by decreasing code complexity and increasing code reuse~(\secref{ana_scale}).


\subsection{Network modules -- software aggregation and modularity}
\label{sec_ana_module}
Packages of the software system reflect in different structural modules within class dependency networks~\cite{SB11s,SB12u}. For instance, visualization classes commonly aggregate into communities of densely connected nodes~\cite{GN02}, whereas different parsers, transformers or plugins often arrange into functional modules~\cite{SB12f} that correspond to (disconnected) groups of nodes with common linkage patterns. Otherwise, clear community structure signifies highly modular structure of the respective software system, while well supported functional modules are related to clear functional roles of the classes within the project~\cite{SB11s,SB12u,SB12f}.

\tblref{packs} compares software packages against network modules identified with \mo\cite{CNM04} and \cp\cite{SB11d,SB11b} community detection approaches, and \mm\cite{NL07} and \gp\cite{SB12u,SB12f} structural module identification algorithms. Analysis reveals that general structural modules including communities and functional modules most accurately model the package structure of the software systems in this~study.

\begin{figure*}[p]
\centering
\includegraphics[width=2.00\columnwidth]{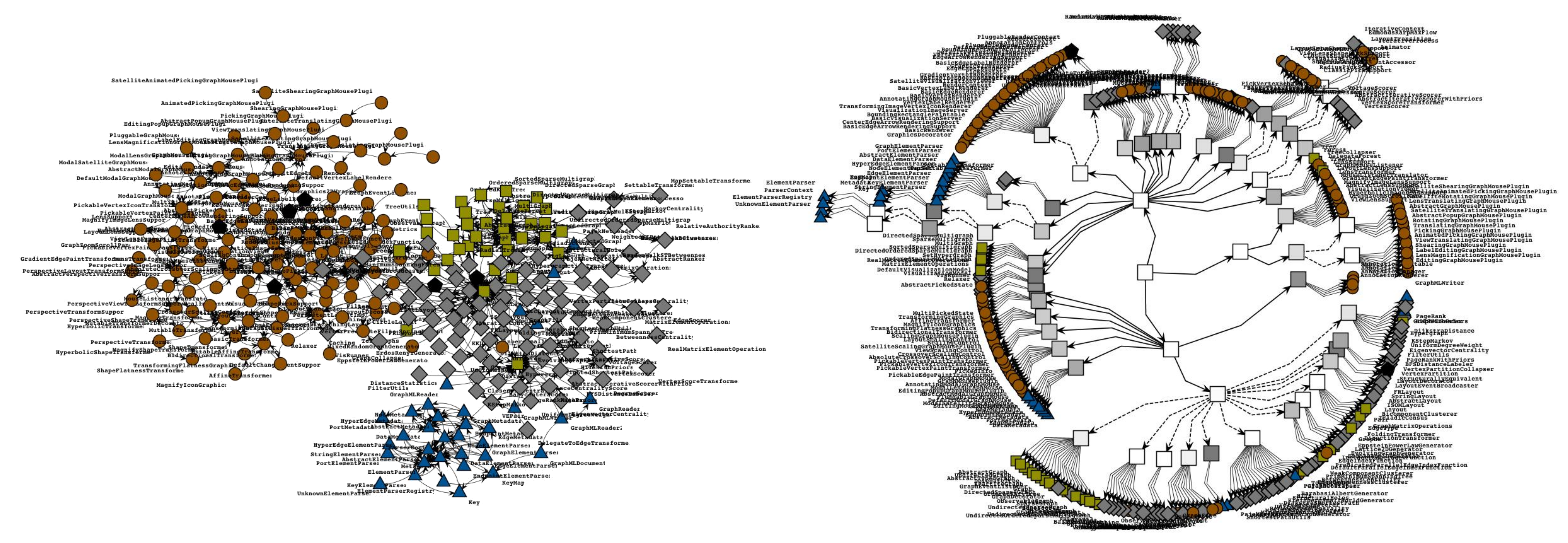}
\caption{\label{fig_abst}(left)~\jng network where node symbols represent high-level packages of JUNG framework: \texttt{visualization} ({\color{red}circles}), \texttt{io} ({\color{blue}triangles}), \texttt{graph} ({\color{green}squares}) and \texttt{algorithms} ({\color{gray}diamonds}). (right)~Hierarchy of structural modules revealed with the algorithm in~\cite{SB12f}.}
\end{figure*}

\begin{figure*}[p]
\centering
\includegraphics[width=2.00\columnwidth]{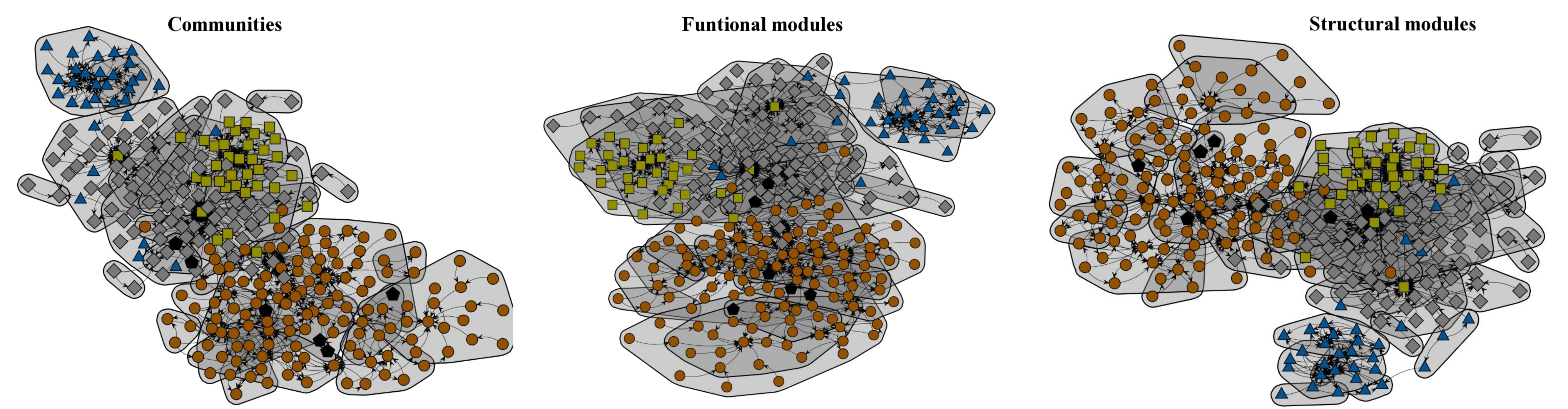}
\caption{\label{fig_ref}(left)~Communities representing highly modular structure of the software system~\cite{SB11d,SB11b}. (middle)~Functional modules that represent highly functional partitioning of the system~\cite{SB12u,SB12f}. (right)~General structural modules conveying modular and functional links (bottom-most level of the hierarchy in~\figref{abst}).}
\end{figure*}

\begin{table*}[p]
\centering
\caption{\label{tbl_packs}Normalized mutual information~\cite{DDDA05} (NMI) between software packages and identified network modules, $\mathrm{NMI}\in[0,1]$. Number of modules is shown in small font.}
\begin{tabular}{|c|l|r|l|r|l|r|l|r|l|}
\hline
\multicolumn{2}{|c}{\rule{0pt}{12pt}Network} & 
\multicolumn{2}{|c}{\mo} & 
\multicolumn{2}{|c}{\cp} & 
\multicolumn{2}{|c}{\mm} & 
\multicolumn{2}{|c|}{\gp}\\\hline
\flm & {\scriptsize $16$} & $0.580$ & {\scriptsize $14$} & $\mathbf{0.609}$ & {\scriptsize $27$} & $0.521$ & {\scriptsize $16$} & $\mathbf{0.610}$ & {\scriptsize $26$} \\\hline
\clt & {\scriptsize $19$} & $0.519$ & {\scriptsize $10$} & $0.473$ & {\scriptsize $20$} & $\mathbf{0.533}$ & {\scriptsize $19$} & $\mathbf{0.530}$ & {\scriptsize $26$} \\\hline
\jng & {\scriptsize $39$} & $0.614$ & {\scriptsize $13$} & $0.650$ & {\scriptsize $30$} & $0.661$ & {\scriptsize $39$} & $\mathbf{0.680}$ & {\scriptsize $41$} \\\hline
\org & {\scriptsize $47$} & $0.503$ & {\scriptsize $11$} & $\mathbf{0.537}$ & {\scriptsize $30$} & $0.378$ & {\scriptsize $39$} & $\mathbf{0.536}$ & {\scriptsize $33$} \\\hline
\wek & {\scriptsize $81$} & $\mathbf{0.558}$ & {\scriptsize $26$} & $0.410$ & {\scriptsize $49$} & $0.430$ & {\scriptsize $63$} & $0.314$ & {\scriptsize $28$} \\\hline
\jvx & {\scriptsize $107$} & $0.704$ & {\scriptsize $59$} & $\mathbf{0.761}$ & {\scriptsize $155$} & $0.392$ & {\scriptsize $89$} & $0.747$ & {\scriptsize $192$} \\\hline
\end{tabular}
\end{table*}

\begin{table*}[p]
\centering
\caption{\label{tbl_ca}Classification accuracy (CA) for software package prediction, $\mathrm{CA}\in[0,1]$. ($l_{\infty}$ is the number of levels of the package hierarchy, whereas $l$ is the average level for a software class. Value under $P_i$ corresponds to CA for the $i$-th level of the hierarchy.)}
\begin{tabular}{|c|r|c|r|r|r|r|r|}
\hline
\multicolumn{1}{|c}{\rule{0pt}{12pt}Network} & 
\multicolumn{1}{|c}{$l$} &
\multicolumn{1}{|c}{$l_{\infty}$} &
\multicolumn{1}{|c}{$P$} &
\multicolumn{1}{|c}{$P_{4}$} & 
\multicolumn{1}{|c}{$P_{3}$} & 
\multicolumn{1}{|c}{$P_{2}$} & 
\multicolumn{1}{|c|}{$P_{1}$}\\\hline
\flm & $2.65$ & $4$ & $\mathbf{0.566}$ & \multicolumn{1}{c|}{$\leftarrow$} & $0.572$ & $\mathit{0.793}$ & $1.000$ \\\hline
\clt & $3.35$ & $4$ & $\mathbf{0.654}$ & \multicolumn{1}{c|}{$\leftarrow$} & $\mathit{0.756}$ & $0.942$ & $1.000$ \\\hline
\jng & $2.97$ & $4$ & $\mathbf{0.617}$ & \multicolumn{1}{c|}{$\leftarrow$} & $0.663$ & $\mathit{0.857}$ & $1.000$ \\\hline
\org & $3.50$ & $7$ & $\mathbf{0.616}$ & $0.616$ & $\mathit{0.714}$ & $0.989$ & $1.000$ \\\hline
\wek & $3.02$ & $6$ & $\mathbf{0.684}$ & $0.692$ & $\mathit{0.736}$ & $0.871$ & $1.000$ \\\hline
\jvx & $3.11$ & $5$ & $\mathbf{0.626}$ & $0.631$ & $\mathit{0.816}$ & $0.982$ & $1.000$ \\\hline
\end{tabular}
\end{table*}


\section{Applications}
\label{sec_appl}
Due to space limitations, the following section only briefly describes different applications of network analysis techniques presented in~\secref{ana}. Future work will focus on a more detailed examination and development of supporting implementations that could be easily applied in practice.

\subsection{Software project abstraction}
\figref{abst}~shows an application of network structural module detection to software project abstraction. One can identify an entire hierarchy of modules that is consistent with the package hierarchy, while also enclosing class dependencies that go beyond packages decided by the developers. Besides better comprehension, revealed hierarchy enables the prediction of dependencies between the classes of a project~\cite{SB12f}.

\begin{table*}[t]
\centering
\caption{\label{tbl_pind}Software project quality indicators presented in the study. For each indicator, we give the range and the expected value of a well designed software system (based on~\secref{ana}).}
\begin{tabular}{|c|c|c|l|}
\hline
\multicolumn{1}{|c}{Quality indicator} & 
\multicolumn{1}{|c}{Expected value} & 
\multicolumn{1}{|c}{Range} & 
\multicolumn{1}{|c|}{Comment}\\\hline
$p_k^{in}$ & $\sim k^{-\gamma_{in}}$ & $\gamma_{in}>1$ & \multirow{2}{*}{High code reusability (\secref{ana_scale}).} \\\cline{1-3}
$k^{in}$ & $\gg 0$ & $\infty$ & \\\hline
$p_k^{out}$ & $\nsim k^{-\gamma_{out}}$ & $\gamma_{out}>1$ & \multirow{2}{*}{Low code complexity (\secref{ana_scale}).} \\\cline{1-3}
$k^{out}$ & $\ll n$ & $\infty$ & \\\hline
$D$ & $\gg 0$, $\ll 1$ & $[0,1]$ & Characteristics of the project domain~\cite{SB12f}. \\\hline
$l-l_{ER}$ & $\leq 0$ & $\infty$ & Well structured and designed project (\secref{ana_world}). \\\hline
$E$ & $\approx 0$ & $[0,1]$ & Low efficiency of information flow (\secref{ana_world}). \\\hline
$\drive$ & $\gg 0$ & $[0,1]$ & Low project controllability (\secref{ana_node}). \\\hline
$\gamma$ & $\ll 3$ & $>1$ & Low project vulnerability, high robustness (\secref{ana_scale}). \\\hline
\end{tabular}
\end{table*}

\begin{table*}[t]
\centering
\caption{\label{tbl_cind}Software class indicators presented in the study. For each indicator, we give the range and the expected value of highly influential, most vulnerable or high complexity classes (based on~\secref{ana}).}
\begin{tabular}{|c|c|c|l|}
\hline
\multicolumn{1}{|c}{\rule{0pt}{12pt}Class indicator} & 
\multicolumn{1}{|c}{\rule{0pt}{12pt}Expected value} & 
\multicolumn{1}{|c}{\rule{0pt}{12pt}Range} & 
\multicolumn{1}{|c|}{\rule{0pt}{12pt}Comment}\\\hline
$\dcent$, $\bcent$ & \multirow{2}{*}{$\gg 0$} & \multirow{2}{*}{$[0,1]$} & Highly influential seed classes (\secref{ana_node}). \\\cline{1-1}\cline{4-4}
$\ccent$ & & & Highly vulnerable seed classes (\secref{ana_node}). \\\hline
$k^{in}$ & \multirow{2}{*}{$\gg 0$} & \multirow{2}{*}{$\infty$} & Highly influential hub classes (\secref{ana_scale}). \\\cline{1-1}\cline{4-4}
$k^{out}$ &  & & High complexity hub classes (\secref{ana_scale}). \\\hline
\end{tabular}
\end{table*}


\subsection{Software packages refactoring}
Network module detection algorithms can also be applied for refactoring of software packages~\cite{SB11s,SB12f}. One can adopt a community detection algorithm to reveal highly modular structure (\figref{ref},~(left)) or a functional module detection algorithm to identify the underlying functional structure (\figref{ref},~(middle)). General structural module detection algorithms partition software classes according to both modular and functional links that are present among the dependencies of the project (\figref{ref},~(right)).


\subsection{Software packages prediction}
\tblref{ca}~shows classification accuracies for the prediction of software packages for the classes of different systems. Let $i$ be a node corresponding to class $A_i$. Package of $A_i$ is then predicted to be the most likely package considering nodes within the same structural module as $i$. The nodes are weighted according to Jaccard similarity~\cite{Jac01}, which is defined as $|\Gamma_i\cap\Gamma_j|/|\Gamma_i\cup\Gamma_j|$, where $j$ is a similar node and $\Gamma_i$ is the neighborhood of node $i$. Structural modules are identified with the algorithm in~\cite{SB12u,SB12f}.

On average, one can predict software packages with $\approx 80\%$ probability for most classes of the systems considered, whereas complete package hierarchy can be precisely identified for over $60\%$ of the software classes (\tblref{ca}).


\subsection{Software quality indicators}
\tblref{pind} and \tblref{cind} show software project and class quality indicators identified in the study. Indicators can be employed to assess project structure and design, code complexity and reusability, controllability and vulnerability, information flow, and other. Due to space limitations, comparison with other approaches for measuring software quality is omitted (e.g., metrics of coupling and cohesion~\cite{SMC99}).


\section{Conclusions}
\label{sec_conc}
The paper conducts a comprehensive study of software networks constructed from Java source code. First, we address macroscopic network properties that are related to structural design of the corresponding software project. Next, we analyze the networks on a microscopic level of nodes, to highlight most influential and vulnerable software classes. Last, we analyze mesoscopic network structural modules and expose their applicability in project refactoring. Among other, we show that software systems are highly vulnerable to processes like bug propagation, however, they are not easily controllable. On the other hand, Java language can be controlled through merely $17\%$ of \texttt{java} namespace. We also identify several network-based quality indicators that can be employed to assess software project design, reusability, robustness, controllability and other. The study thus exposes network analysis as a prominent set of tools for software systems engineering.



\section{Acknowledgments}
This work has been supported by the Slovene Research Agency ARRS within Research Program No. P2-0359.


\bibliographystyle{abbrv}

\begin{thebibliography}{10}

\bibitem{AMO93}
R.~K. Ahuja, T.~L. Magnanti, and J.~B. Orlin.
\newblock {\em Network flows: Theory, algorithms, and applications}.
\newblock {Prentice-Hall}, Upper Saddle River, {NJ}, 1993.

\bibitem{AJB00}
R.~Albert, H.~Jeong, and A.~L. Barabasi.
\newblock Error and attack tolerance of complex networks.
\newblock {\em Nature}, 406(6794):378--382, 2000.

\bibitem{BA99}
A.~L. Barab\'{a}si and R.~Albert.
\newblock Emergence of scaling in random networks.
\newblock {\em Science}, 286(5439):509--512, 1999.

\bibitem{BFNRSVMT06}
G.~Baxter, M.~Frean, J.~Noble, M.~Rickerby, H.~Smith, M.~Visser, H.~Melton, and
  E.~Tempero.
\newblock Understanding the shape of java software.
\newblock In {\em Proceedings of the {ACM} International Conference on
  {Object-Oriented} Programming, Systems, Languages, and Applications}, pages
  397--412, 2006.

\bibitem{Bei90}
B.~Beizer.
\newblock {\em Software testing techniques}.
\newblock Van Nostrand Reinhold Co., New York, {NY}, {USA}, 1990.

\bibitem{CY09}
K.~Cai and B.~Yin.
\newblock Software execution processes as an evolving complex network.
\newblock {\em Information Sciences}, 179(12):1903--1928, 2009.

\bibitem{CGP00}
E.~M. Clarke, O.~Grumberg, and D.~Peled.
\newblock {\em Model checking}.
\newblock {MIT} Press, 2000.

\bibitem{CNM04}
A.~Clauset, M.~E.~J. Newman, and C.~Moore.
\newblock Finding community structure in very large networks.
\newblock {\em Physical Review E}, 70(6):066111, 2004.

\bibitem{CMPS07}
G.~Concas, M.~Marchesi, S.~Pinna, and N.~Serra.
\newblock Power-laws in a large object-oriented software system.
\newblock {\em {IEEE} Transactions on Software Engineering}, 33(10):687--708,
  2007.

\bibitem{DDDA05}
L.~Danon, A.~{D\'{i}az-Guilera}, J.~Duch, and A.~Arenas.
\newblock Comparing community structure identification.
\newblock {\em Journal of Statistical Mechanics: Theory and Experiment},
  P09008, 2005.

\bibitem{Ege11a}
M.~Egerstedt.
\newblock Complex networks: Degrees of control.
\newblock {\em Nature}, 473(7346):158--159, 2011.

\bibitem{ER59}
P.~Erd\H{o}s and A.~R\'{e}nyi.
\newblock On random graphs i.
\newblock {\em Publicationes Mathematicae Debrecen}, 6:290--297, 1959.

\bibitem{Fre77}
L.~Freeman.
\newblock A set of measures of centrality based on betweenness.
\newblock {\em Sociometry}, 40(1):35--41, 1977.

\bibitem{Fre79}
L.~C. Freeman.
\newblock Centrality in social networks: Conceptual clarification.
\newblock {\em Social Networks}, 1(3):215--239, 1979.

\bibitem{GBSH12}
J.~Gao, S.~V. Buldyrev, H.~E. Stanley, and S.~Havlin.
\newblock Networks formed from interdependent networks.
\newblock {\em Nature Physics}, 8(1):40--48, 2012.

\bibitem{GN02}
M.~Girvan and M.~E.~J. Newman.
\newblock Community structure in social and biological networks.
\newblock {\em Proceedings of the National Academy of Sciences of United States
  of America}, 99(12):7821--7826, 2002.

\bibitem{HCK06}
D.~{Hyland-Wood}, D.~Carrington, and S.~Kaplan.
\newblock Scale-free nature of java software package, class and method
  collaboration graphs.
\newblock In {\em Proceedings of the International Symposium on Empirical
  Software Engineering}, pages 1--10, 2006.

\bibitem{Jac01}
P.~Jaccard.
\newblock \'{E}tude comparative de la distribution florale dans une portion des
  alpes et des jura.
\newblock {\em Bulletin del la Soci\'{e}t\'{e} Vaudoise des Sciences
  Naturelles}, 37:547--579, 1901.

\bibitem{Koh09}
G.~A. Kohring.
\newblock Complex dependencies in large software systems.
\newblock {\em Advances in Complex Systems}, 12(6):565--581, 2009.

\bibitem{KDG07}
A.~Kuhn, S.~Ducasse, and T.~G\^{i}rba.
\newblock Semantic clustering: Identifying topics in source code.
\newblock {\em Information and Software Technology}, 49(3):230--243, 2007.

\bibitem{LW04}
N.~{LaBelle} and E.~Wallingford.
\newblock Inter-package dependency networks in open-source software.
\newblock {\em e-print {arXiv:cs/0411096v1}}, 2004.

\bibitem{Lm01}
V.~Latora and M.~Marchiori.
\newblock Efficient behavior of small-world networks.
\newblock {\em Physical Review Letters}, 87(19):198701, 2001.

\bibitem{LSB11b}
Y.~Liu, J.~Slotine, and A.~Barabasi.
\newblock Controllability of complex networks.
\newblock {\em Nature}, 473(7346):167--173, 2011.

\bibitem{LH07}
A.~Lombardi and M.~H\"{o}rnquist.
\newblock Controllability analysis of networks.
\newblock {\em Physical Review E}, 75(5):056110, 2007.

\bibitem{Mil67}
S.~Milgram.
\newblock The small world problem.
\newblock {\em Psychology Today}, 1(1):60--67, 1967.

\bibitem{MRMPO10}
P.~J. Mucha, T.~Richardson, K.~Macon, M.~A. Porter, and J.~Onnela.
\newblock Community structure in time-dependent, multiscale, and multiplex
  networks.
\newblock {\em Science}, 328(5980):876--878, 2010.

\bibitem{Mye03}
C.~R. Myers.
\newblock Software systems as complex networks: Structure, function, and
  evolvability of software collaboration graphs.
\newblock {\em Physical Review E}, 68(2), 2003.

\bibitem{NGS08}
G.~M. Narayan, K.~Gopinath, and V.~Sridhar.
\newblock Structure and interpretation of computer programs.
\newblock In {\em Proceedings of the {IEEE} International Symposium on
  Theoretical Aspects of Software Engineering}, 2008.

\bibitem{NL07}
M.~E.~J. Newman and E.~A. Leicht.
\newblock Mixture models and exploratory analysis in networks.
\newblock {\em Proceedings of the National Academy of Sciences of United States
  of America}, 104(23):9564, 2007.

\bibitem{PV01}
R.~{Pastor-Satorras} and A.~Vespignani.
\newblock Epidemic spreading in scale-free networks.
\newblock {\em Physical Review Letters}, 86(14):3200--3203, 2001.

\bibitem{PNFB05}
A.~Potanin, J.~Noble, M.~Frean, and R.~Biddle.
\newblock Scale-free geometry in {OO} programs.
\newblock {\em Communications of the {ACM}}, 48(5):99--103, 2005.

\bibitem{Sin11}
S.~Sinha.
\newblock Few and far between.
\newblock {\em Physics}, 4:81, 2011.

\bibitem{SV05}
S.~N. Soffer and A.~V\'{a}zquez.
\newblock Network clustering coefficient without degree-correlation biases.
\newblock {\em Physical Review E}, 71(5):057101, 2005.

\bibitem{SMC99}
W.~P. Stevens, G.~J. Myers, and L.~L. Constantive.
\newblock Structured design.
\newblock {\em {IBM} Systems Journal}, 38(2):231--256, 1999.

\bibitem{VCS02}
S.~Valverde, R.~F. Cancho, and R.~V. Sol\'{e}.
\newblock Scale-free networks from optimal design.
\newblock {\em Europhysics Letters}, 60(4):512, 2002.

\bibitem{VS05a}
S.~Valverde and R.~V. Sol\'{e}.
\newblock Logarithmic growth dynamics in software networks.
\newblock {\em Europhysics Letters}, 72(5):858--864, 2005.

\bibitem{VS05b}
S.~Valverde and R.~V. Sol\'{e}.
\newblock Network motifs in computational graphs: A case study in software
  architecture.
\newblock {\em Physical Review E}, 72(2):026107, 2005.

\bibitem{VS07}
S.~Valverde and R.~V. Sol\'{e}.
\newblock Hierarchical small worlds in software architecture.
\newblock {\em Dynamics of Continuous, Discrete and Impulsive Systems - Series
  B}, 14:1--11, 2007.

\bibitem{SB11s}
L.~\v{S}ubelj and M.~Bajec.
\newblock Community structure of complex software systems: Analysis and
  applications.
\newblock {\em Physica A: Statistical Mechanics and its Applications},
  390(16):2968--2975, 2011.

\bibitem{SB11g}
L.~\v{S}ubelj and M.~Bajec.
\newblock Generalized network community detection.
\newblock In {\em Proceedings of the {ECML} {PKDD} Workshop on Finding Patterns
  of Human Behaviors in Network and Mobility Data}, pages 66--84, Athens,
  Greece, 2011.

\bibitem{SB11b}
L.~\v{S}ubelj and M.~Bajec.
\newblock Robust network community detection using balanced propagation.
\newblock {\em European Physical Journal B}, 81(3):353--362, 2011.

\bibitem{SB11d}
L.~\v{S}ubelj and M.~Bajec.
\newblock Unfolding communities in large complex networks: Combining defensive
  and offensive label propagation for core extraction.
\newblock {\em Physical Review E}, 83(3):036103, 2011.

\bibitem{SB12f}
L.~\v{S}ubelj and M.~Bajec.
\newblock Clustering assortativity, communities and functional modules in
  real-world networks.
\newblock {\em e-print {arXiv:12023188v1}}, 2012.

\bibitem{SB12u}
L.~\v{S}ubelj and M.~Bajec.
\newblock Ubiquitousness of link-density and link-pattern communities in
  real-world networks.
\newblock {\em European Physical Journal B}, 85(1):32, 2012.

\bibitem{WL10a}
J.~Wang and Y.~Liu.
\newblock Modeling software faults propagation.
\newblock {\em Europhysics Letters}, 92(6):60009, 2010.

\bibitem{WS98}
D.~J. Watts and S.~H. Strogatz.
\newblock Collective dynamics of 'small-world' networks.
\newblock {\em Nature}, 393(6684):440--442, 1998.

\bibitem{WKD07}
L.~Wen, D.~Kirk, and R.~G. Dromey.
\newblock Software systems as complex networks.
\newblock In {\em Proceedings of the {IEEE} International Conference on
  Cognitive Informatics}, pages 106--115, 2007.

\end{thebibliography}

\end{document}